\documentclass[conference]{IEEEtran}
\IEEEoverridecommandlockouts
\usepackage{cite}
\usepackage{amsmath,amssymb,amsfonts}
\usepackage{algorithmic}
\usepackage{graphicx}
\usepackage{ragged2e}
\usepackage{adjustbox}
\usepackage{url}
\usepackage[most]{tcolorbox}
\usepackage{float}
\graphicspath{ {./figures/} } 
\usepackage{textcomp}
\usepackage{tikz}
\usetikzlibrary{shapes.multipart, positioning, arrows.meta, fit, calc}
\usepackage[colorlinks=false, hidelinks]{hyperref}
\usepackage{xcolor}
\usepackage{caption}
\def\BibTeX{{\rm B\kern-.05em{\sc i\kern-.025em b}\kern-.08em
    T\kern-.1667em\lower.7ex\hbox{E}\kern-.125emX}}
\begin{document}

\title{MultiFuzz: A Dense Retrieval-based Multi-Agent System for Network Protocol Fuzzing \\
{\footnotesize \textsuperscript{}}}

\author{
\IEEEauthorblockN{{\large Youssef Maklad\IEEEauthorrefmark{1},
Fares Wael\IEEEauthorrefmark{1},
Ali Hamdi\IEEEauthorrefmark{1},
Wael Elsersy\IEEEauthorrefmark{1},
Khaled Shaban\IEEEauthorrefmark{2}}}

\IEEEauthorblockA{\IEEEauthorrefmark{1}\large\textit{Dept. of Computer Science, MSA University, Giza, Egypt} \\
\{youssef.mohamed88, fares.wael, ahamdi, wfarouk\}@msa.edu.eg}

\IEEEauthorblockA{\IEEEauthorrefmark{2}\large\textit{Dept. of Computer Science, Qatar University, Doha, Qatar} \\
khaled.shaban@qu.edu.qa}
}

\maketitle

\begin{abstract}
Traditional protocol fuzzing techniques, such as those employed by AFL-based systems, often lack effectiveness due to a limited semantic understanding of complex protocol grammars and rigid seed mutation strategies. Recent works, such as ChatAFL, have integrated Large Language Models (LLMs) to guide protocol fuzzing and address these limitations, pushing protocol fuzzers to wider exploration of the protocol state space. But ChatAFL still faces issues like unreliable output, LLM hallucinations, and assumptions of LLM knowledge about protocol specifications. This paper introduces MultiFuzz, a novel dense retrieval-based multi-agent system designed to overcome these limitations by integrating semantic-aware context retrieval, specialized agents, and structured tool-assisted reasoning. MultiFuzz utilizes agentic chunks of protocol documentation (RFC Documents) to build embeddings in a vector database for a retrieval-augmented generation (RAG) pipeline, enabling agents to generate more reliable and structured outputs, enhancing the fuzzer in mutating protocol messages with enhanced state coverage and adherence to syntactic constraints. The framework decomposes the fuzzing process into modular groups of agents that collaborate through chain-of-thought reasoning to dynamically adapt fuzzing strategies based on the retrieved contextual knowledge. Experimental evaluations on the Real-Time Streaming Protocol (RTSP) demonstrate that MultiFuzz significantly improves branch coverage and explores deeper protocol states and transitions over state-of-the-art (SOTA) fuzzers such as NSFuzz, AFLNet, and ChatAFL. By combining dense retrieval, agentic coordination, and language model reasoning, MultiFuzz establishes a new paradigm in autonomous protocol fuzzing, offering a scalable and extensible foundation for future research in intelligent agentic-based fuzzing systems.
\end{abstract}

% \vspace{0.12cm}
    

% ------------- START Introduction --------------------
\section{Introduction}
Network protocols form the backbone of modern communication systems, yet remain vulnerable to many flaws that can compromise entire infrastructures' security. Protocol fuzzing has long been an effective technique for uncovering these vulnerabilities through automated test generation, and it has long been recognized as an effective technique for uncovering these software vulnerabilities \cite{fuzzing-book}. As network services grow in complexity and scale, the importance of discovering implementation flaws, especially in stateful protocols with finite-state machines (FSMs), increases. Network protocol fuzzing attempts to systematically test protocol implementations by generating malformed, unexpected, or semi-valid protocol messages to identify anomalous behavior. However, traditional fuzzing methods often struggle with unique protocol challenges, such as handling complex grammar formats, managing deep-protocol state transitions, and maintaining valid session semantics across multi-packet interactions \cite{NPF-Survey-New1}.

Recent research highlight multiple directions in the advancement of protocol fuzzing, including state-aware input generation and automated reverse engineering of undocumented protocols \cite{survey-protocol-reverse-engineering, state-selection-algos-imp}. Despite these developments, achieving high coverage and deeper state exploration remains difficult, particularly for closed-source or proprietary protocols. This has motivated the integration of more intelligent components into the fuzzing loop.

The rise of LLMs has opened new avenues for automating traditionally manual tasks in software engineering. LLMs have demonstrated strong capabilities in reasoning, code understanding, and program synthesis \cite{eval-llms-trained-on-code, lmsCanSolve}. Their potential in security applications, including fuzzing, has begun to evolve. Studies show that LLMs can infer message grammars, generate valid input sequences, and even simulate stateful behavior without access to source code \cite{LLM-NPF-Survey, GenAIForCybersecurity}. Recent works such as ChatAFL introduced LLM-guided protocol fuzzing, resulting in improved protocol state coverage \cite{chatafl}. These developments highlight LLMs as promising assistants for fuzzing complex, stateful, and security-critical systems.

In this work, we present \textit{MultiFuzz}, a multi-agent system built on top of ChatAFL, designed to enhance network protocol fuzzing by unleashing the capabilities of LLMs and dense retrieval \cite{denseXret}. Inspired by recent advancements in retrieval-augmented generation and ReAct-based chain-of-thought reasoning \cite{rag, react}, \textit{MultiFuzz} is structured around collaborative agents, each responsible for a specific phase of the ChatAFL fuzzing pipeline. Unlike traditional fuzzers or single LLM approaches, \textit{MultiFuzz} orchestrates tool-augmented agents to maintain semantic context support and protocol-specific inference.

The main contributions of this work are as follows:

\begin{itemize}
    \item We propose \textit{MultiFuzz}, a multi-agent system for protocol fuzzing, integrated with the ChatAFL framework, where each group of agents is dedicated to a specific subtask and enhanced with tool integration, vector database context awareness, and CVE-driven vulnerability knowledge.

     \item We introduce an agentic-chunking method and embedding strategy for protocol RFC documents, enabling semantic indexing of protocol knowledge for agent use.
    
    \item We integrate dense retrieval into the agent reasoning process to maintain a protocol-aware context and guide more effective fuzzing actions.
    
    \item We evaluate \textit{MultiFuzz} on stateful protocol targets, and the results demonstrate improvements in branch coverage, number of states explored, and state transitions compared to SOTA fuzzers such as NSFuzz, AFLNet, and ChatAFL.
\end{itemize}

Through \textit{MultiFuzz}, we aim to bridge the gap between protocol-aware fuzzing needs and the generative coordination capabilities of modern agentic-AI architecture, allowing more intelligent and effective protocol fuzzing.

% The paper is organized as follows: section \ref{rw} outlines the related work. Section \ref{bg} provides background on FSM extraction techniques and LLMs. Section \ref{methodology} details our methodology. Section \ref{ex-des} discusses the research questions and the evaluation of the framework, while section \ref{res} presents the experimental results. Finally, section \ref{conc} concludes the paper and suggests potential directions for future work.

The paper is structured as follows: Section \ref{bg} introduces background on protocol fuzzing, LLMs, and multi-agent systems. Section \ref{rw} reviews the related work. Section \ref{methodology} explains our methodology. Section \ref{ex-des} explores the research questions, experiments' setup, and evaluation metrics. Section \ref{res} highlights the experimental results. Finally, section \ref{conc} concludes the paper and suggests potential directions for future works.
% ------------- END Introduction --------------------

% ------------- START Background --------------------
\section{Background}\label{bg}
The following subsections review key concepts and recent developments in network protocol fuzzing, LLMs, and multi-agent systems, providing background information for their integration in modern frameworks.

\subsection{Network Protocol Fuzzing}
% Protocol fuzzing is a specialized security-testing technique that targets the finite-state behavior of communication protocols by injecting crafted or mutated packets to uncover flaws. It relies on intelligently generating seed inputs that exercise different protocol states, since exploring deep protocol behaviors often exposes hidden vulnerabilities. Network protocol fuzzing focuses on testing stateful network services (e.g. TCP/IP, FTP, SMTP, or custom protocols) by feeding sequences of packets through the protocol’s finite-state machine (FSM) in the server under test (SUT). The goal is to traverse unusual protocol paths and trigger implementation bugs or security flaws. Traditional fuzzers such as AFL-based systems \cite{afl, afl++, aflnet} often rely on random bit-flipping or fixed grammar-based inputs, which may fail to reach deep protocol states.

Protocol fuzzing is a specialized security-testing technique that targets the finite-state behavior of communication protocols by injecting crafted or mutated packets to uncover flaws. It relies on intelligently generating seed inputs that exercise different protocol states, since exploring deep protocol behaviors often exposes hidden vulnerabilities. Network protocol fuzzing focuses on testing stateful network services by feeding packet sequences through the protocol’s FSM in the server under test (SUT). The goal is to traverse unusual protocol paths and trigger implementation bugs or security flaws. Fuzzers can be broadly classified by their test case generation strategy. Mutation-based approaches modify existing valid packets using bit-flipping, arithmetic, block-level, or dictionary-based transformations \cite{artOfFuzzing}. In contrast, generation-based approaches synthesize packets from protocol specifications or templates. While mutation methods may struggle with diverse data types and protocol constraints, generation-based methods often face difficulties in acquiring or modeling accurate protocol specifications. Additionally, based on the level of knowledge about the target system, fuzzing can be categorized into blackbox, whitebox, and graybox approaches. Blackbox fuzzers operate without internal knowledge of the protocol implementation, relying solely on input/output observation. White-box fuzzers analyze the source code to guide test case generation, while gray-box fuzzers use lightweight instrumentation as code coverage feedback to guide mutations more effectively. In practice, graybox fuzzing offers a balanced trade-off and is widely used due to limited access to source code in real-world protocol implementations \cite{NPF-Survey}.

\subsection{Large Language Models}
LLMs have recently demonstrated powerful capabilities in generating and reasoning over complex inputs, opening new opportunities for automation in domains like software testing and cybersecurity \cite{LLM-NPF-Survey, GenAIForCybersecurity}. They are deep transformer-based neural networks \cite{attention-is-all-you-need}, trained on massive text corpora, enabling them to generate coherent language and perform complex reasoning \cite{react}. Their rich knowledge and generative power have been harnessed in many domains. In cybersecurity, LLMs have shown remarkable utility. For example, ChatPhishDetector uses an LLM to detect phishing websites \cite{chatphishdetector}. Maklad. Y. et al demonstrated how LLMs, enhanced by RAG and chain-of-thought reasoning, can be used to evaluate seed enrichment tasks and network packet generation \cite{ehna}. SeedMind explored the use of LLMs for building fuzzing seed generators \cite{seedmind}. Codamosa highlights the use of LLMs in overcoming coverage plateaus in test generation \cite{codamosa}. LLMs have also been integrated into automation workflows such as Robotic Process Automation (RPA) and OCR \cite{erpa, lmrpa, lmvrpa}. These results highlight that LLMs, when incorporated intelligently, can enhance the accuracy and efficiency of automation systems. 

\subsection{Multi-Agent Systems}
Multi-agent systems consist of multiple autonomous entities (agents) that interact and collaborate to solve tasks. By harnessing the diverse capabilities and roles of individual agents, multi-agent systems can tackle complex problems more effectively than a single agent could \cite{llmMAS}. For example, agents might divide a workflow so that some gather information, others perform reasoning, and yet others execute actions. However, orchestrating agents also introduces challenges like optimal task allocation, sharing complex context information, and memory management, which become critical in LLM-based multi-agent architectures. In security applications, multi-agent LLM architectures have begun to emerge. A recent effort has introduced PentestAgent \cite{pentestagent}, a framework in which multiple agents collaborate to automate penetration testing and vulnerability analysis. This demonstrates how multi-agent systems can decompose a complex task (like pentesting) into subtasks handled by specialized LLM agents, improving overall efficiency.
% ------------- END Background --------------------

\begin{figure*}[t]
    \centering
    \includegraphics[scale=0.5]{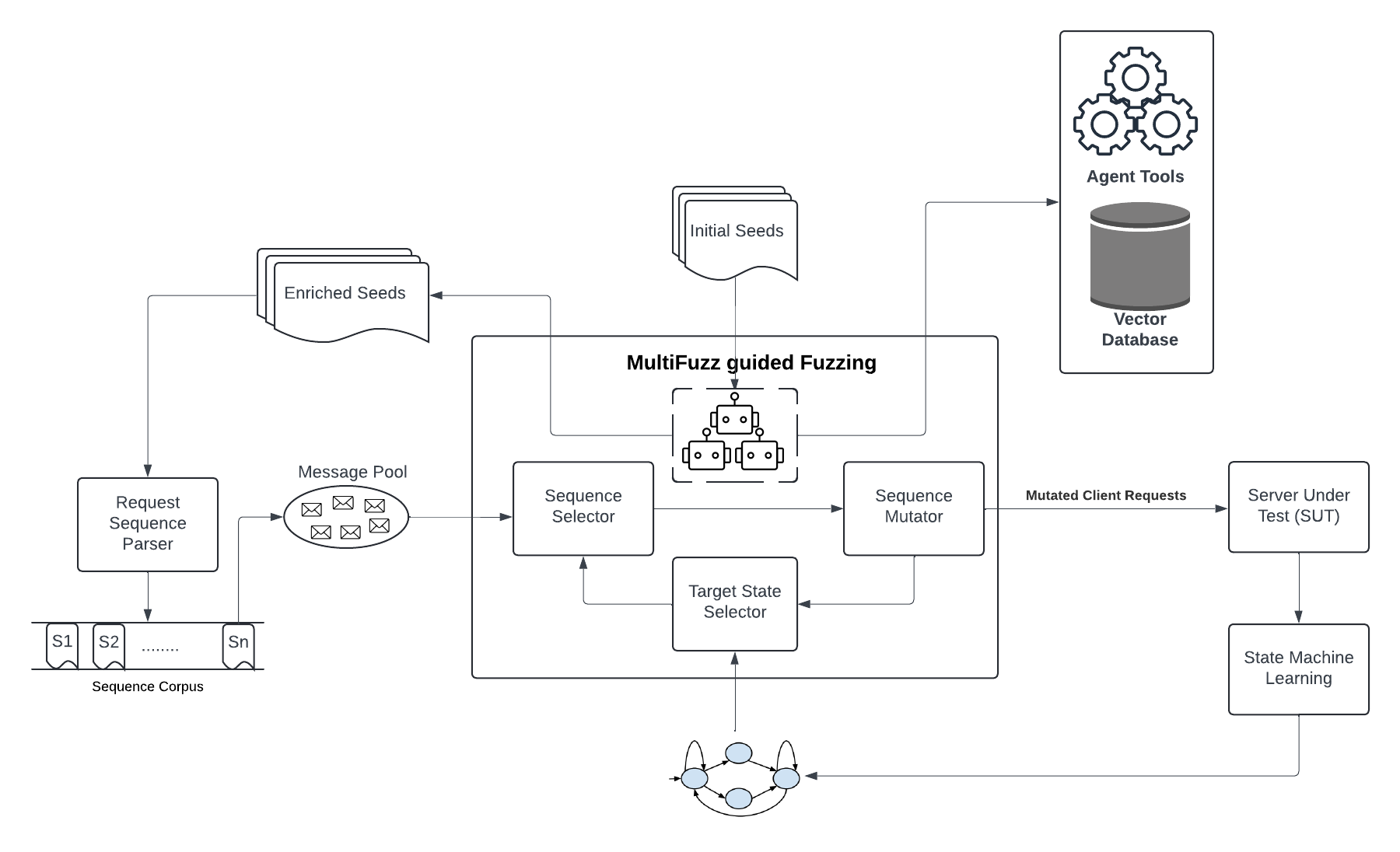}
    \caption{\centering \small Fig. 1. High-Level System Architecture of the \textit{MultiFuzz} Framework, based on AFLNet and ChatAFL}
    \label{fig:System_Architecture}
\end{figure*}

% ------------- START Related Work --------------------
\section{Related Work}\label{rw}
\subsection{Protocol Fuzzing}
Fuzzing has proven to be one of the most effective techniques for vulnerability discovery. Protocol fuzzing, in particular, poses unique challenges due to its reliance on structured formats and stateful interactions. Existing techniques outlined earlier in section \ref{bg}, which are blackbox, whitebox, and graybox fuzzing approaches, offer different trade-offs between scalability, precision, and required prior knowledge of the target protocol.

\subsubsection{\textbf{Blackbox Fuzzing}}
Blackbox fuzzers operate without any internal knowledge of the target and typically rely on traffic observation or mutation of recorded protocol messages. Tools like SPIKE and Peach exemplify early blackbox efforts, relying on manual specification of protocol structures \cite{peach}. PULSAR and BBuzz extract message formats from captured network traffic using protocol reverse engineering techniques \cite{bbuzz, pulsar}. These methods are simple to deploy but struggle with exploring deeper protocol states, often failing to maintain session validity across multi-message interactions.

\subsubsection{\textbf{Whitebox Fuzzing}}
Whitebox fuzzers leverage full access to source code or binaries to systematically explore execution paths. Symbolic execution and taint analysis are commonly used to analyze input-dependent behaviors. Polar combines static analysis with dynamic taint tracking to extract input-related conditions from protocol code \cite{NPF-Survey}. While these techniques offer fine-grained insight and deeper coverage, their scalability is limited by path explosion and instrumentation overhead. Whitebox fuzzers are less commonly used in network protocol contexts due to the complexity of protocol stacks and message interleaving.

\subsubsection{\textbf{Graybox Fuzzing}}
Graybox fuzzers balance insight and scalability by utilizing lightweight instrumentation to guide input mutations. AFL and its extensibles, like AFL++ and AFLNet, employ coverage feedback to direct test-case generation \cite{afl, afl++, aflnet}. AFLNet extends AFL to stateful network protocols by using response codes to infer protocol state transitions. NSFuzz further improves graybox fuzzing by extracting program variables associated with state changes to better synchronize test inputs with protocol logic \cite{nsfuzz}. These tools have proven effective on real-world network services and are the foundation for many modern protocol fuzzers.

\subsection{Large Language Model-assisted Fuzzing}

% \vspace{-1.8cm}

Recent research has explored the integration of LLMs into fuzzing pipelines. These models offer the ability to generate syntactically correct and semantically meaningful inputs by leveraging knowledge learned during pre-training. ChatFuzz uses OpenAI's ChatGPT to mutate existing seed inputs, resulting in improved edge coverage compared to AFL++ \cite{chatfuzz}. ChatAFL constructs message grammars and predicts the next protocol message using GPT models, achieving significant gains in state and code coverage over AFLNet and NSFuzz \cite{chatafl}. MSFuzz extracts abstract syntax trees from the protocol source code via LLMs to guide syntax-aware mutation \cite{msfuzz}. TitanFuzz and FuzzGPT show that LLMs can function as zero-shot fuzzers for deep learning libraries by generating edge case inputs and exploiting rare model behaviors without instrumentation or prior seeds \cite{titanfuzz, fuzzgpt}. These works demonstrate that LLMs can serve as powerful assistants in automating grammar extraction, seed generation, and mutation strategies for protocol fuzzing.
% ------------- END Related Work --------------------

% ------------- START Methodology --------------------
\section{Methodology}\label{methodology}
This section details the design of the \textit{MultiFuzz} framework and its agent-based workflows. The \textit{MultiFuzz} framework APIs are integrated in the ChatAFL framework, on top of the AFLNet architecture. The whole system architecture can be shown in Figure 1. \textit{MultiFuzz} is structured around three specialized crews of agents: the Grammar Extraction Crew, the Seed Enrichment Crew, and the Coverage Plateau Crew. Each crew operates over a shared semantic context retrieved at inference by a common dense retrieval agent, which retrieves the agentic chunked embeddings in the vector store. The workflow begins with preprocessing protocol RFCs, then transforms the content into propositional transformation, followed by agentic chunking, and finally, collaborative agent reasoning at inference.

\vspace{0.25cm}

{\tiny
\begin{tcolorbox}[enhanced,
  sharp corners=south,
  colframe=black!50!black,
  colback=gray!10,
  coltitle=white,
  boxrule=0.6pt,
  titlerule=0.8pt,
  title=System Prompt for Propositional Transformation,
  fonttitle=\scriptsize,
  fontupper=\scriptsize,
  label={box:prop_system_prompt},
]

Decompose the RFC documents' content given into clear and simple text propositions, ensuring they are interpretable out of context.
\vspace{0.05cm}

\textbf{Rules to follow}:
\vspace{0.05cm}

\begin{enumerate}
    \item Split compound sentences into simple sentences. Maintain the original phrasing whenever possible. \vspace{0.05cm}

    \item For any named entity with descriptive information, separate this information into its own distinct proposition. \vspace{0.05cm}

    \item Decontextualize propositions by adding necessary modifiers and replacing pronouns (e.g., "it", "they", "this") with the corresponding full entities. \vspace{0.05cm}

    \item Preserve the structure and formatting of any network packet example, protocol message, or code snippet. Do not summarize them. \vspace{0.05cm}

    \item Present the results as a list of strings formatted in JSON.
\vspace{0.05cm}
\end{enumerate}

\textbf{Example}:
\vspace{0.1cm}

\textit{Input:} "The DESCRIBE method retrieves the description of a media object, it accepts application/sdp..."
\vspace{0.05cm}

\textit{Expected Output:}
\vspace{0.1cm}

\{"sentences": 
["The DESCRIBE method retrieves the description...", "The DESCRIBE method accepts application/sdp..."]

\}
\end{tcolorbox}
}
\begin{center}
    \small Fig. 2. System prompt used for propositional transformation of filtered RFC documents
\end{center}

% \section{Methodology}\label{methodology}
% This section details the design of the \textit{MultiFuzz} framework and its agent-based workflows. The \textit{MultiFuzz} framework APIs are integrated in the ChatAFL framework, on top of the AFLNet architecture. The whole system architecture can be shown in figure \ref{fig:System_Architecture}. \textit{MultiFuzz} is structured around three specialized crews of agents: the Grammar Extraction Crew, the Seed Enrichment Crew, and the Coverage Plateau Crew. Each crew operates over a shared semantic context retrieved at inference by a common dense retrieval agent, which retrieves the agentic chunked embeddings in the vector store. The workflow begins with preprocessing protocol RFCs, then transforms the content into propositional transformation, followed by agentic chunking, and finally, collaborative agent reasoning.

% \vspace{0.0cm}

{\tiny
\begin{tcolorbox}[
  enhanced,
  sharp corners=south,
  colframe=blue!50!black,
  colback=gray!10,
  coltitle=white,
  boxrule=0.6pt,
  titlerule=0.8pt,
  title=Sample Document Chunk Post Agentic Chunking,
  fonttitle=\scriptsize,
  fontupper=\scriptsize,
  label={box:sample_chunk},
]
\hspace{-0.35cm}\{

\textbf{"\textless chunk\_id\textgreater"}: \{ \\
  \textbf{"title"}: "RTSP Streaming Protocols and Resource Management", \\
  \textbf{"summary"}: "This chunk contains information about RTSP controls in streaming protocols, emphasizing session management, RTSP URL semantics, and transmission methods, while including examples and method functionalities.", \\
  \textbf{"propositions"}: [ \\
    "SETUP starts an RTSP session.", \\
    "PLAY starts data transmission on a stream allocated via SETUP.", \\
    "PAUSE does not free server resources.", \\
    "TEARDOWN causes the RTSP session to cease to exist on the server.", \\
    "RTSP methods that contribute to state use the Session header field.", \\
    "The 'rtsp' scheme requires that commands are issued via a reliable protocol, specifically TCP.", \\
    "Lines in RTSP messages are terminated by CRLF.", \\
    "RTSP methods are idempotent unless otherwise noted.", \\
    "For the scheme 'rtsp', a persistent connection is assumed.", \\
    "...", \\
  ] \\
\hspace*{-0.25cm}\} \\
\hspace*{-0.35cm}\}
\end{tcolorbox}
}
\begin{center}
    \small Fig. 3. Sample document chunk after the agentic chunking phase of text propositions. %\vspace{0.025cm}
\end{center}

\subsection{RFC Documents Preprocessing}
We first process the RFC documents, where each RFC is manually segmented into paragraphs and then passes through a series of filters to extract technical sections, including stateful interactions, command formats, and response rules. We define an RFC as a sequence of paragraphs $R = \{r_1, r_2, ..., r_n\}$. A semantic classifier $f_{\text{filter}}$ maps each paragraph to a boolean label:
\begin{equation}
    f_{\text{filter}}(r_i) = \begin{cases}
        1 & \text{if } r_i \text{ is protocol-relevant} \\
        0 & \text{otherwise}
    \end{cases}
\end{equation}

Only filtered paragraphs $R' = \{r_i \in R\ |\ f_{\text{filter}}(r_i) = 1\}$ are retained for downstream chunking and proposition extraction.

\subsection{Propositional Transformation}
Once the RFC content has been semantically filtered and structured into coherent sections using specific delimiters (\texttt{\#\#\#}, \texttt{---}, \texttt{@@@}), the next step in the pipeline is to transform these technical paragraphs into interpretable, context-independent atomic propositions. To perform this transformation, we constructed an LLM-powered pipeline. Each filtered section is first processed using a carefully designed system prompt, shown in Figure 2. The prompt is executed using the \textit{gpt-4o-mini} model with structured output enforced by a \texttt{JSON} schema. Each chunk of the RFC document is passed through this pipeline, producing a list of minimal, decontextualized statements that accurately capture the semantics of the protocol specification. Formally, for a given input chunk $C_i \in \mathcal{C}$ where $\mathcal{C}$ is the set of smart RFC chunks, the transformation function $T$ produces:

\[
T(C_i) = \{p_1, p_2, \dots, p_k\}, \quad \text{where each } p_j \in \mathbb{P}
\]

Here, $\mathbb{P}$ denotes the proposition space containing linguistically simple, context-independent units of meaning. As a result, each paragraph $C_i$ is mapped to a finite set of logically coherent propositions, and the global proposition set $\mathbb{P}$ becomes the knowledge substrate for subsequent dense retrieval and crew-based inference modules. In our experiments on RFC-2326 (RTSP), this step yielded 445 unique and precise propositions.

\begin{figure*}[ht]
\centering
\begin{tikzpicture}[
scale=0.7,
  agent/.style={rectangle, rounded corners=2pt, draw=black, thick, minimum width=2.8cm, minimum height=0.75cm, text centered, font=\scriptsize, fill=blue!5},
  tool/.style={rectangle, draw=black, dashed, thick, minimum width=2.8cm, minimum height=0.6cm, text centered, font=\scriptsize, fill=gray!10},
  crew/.style={draw=black, rounded corners=6pt, thick, inner sep=4pt, font=\bfseries\scriptsize, fill=gray!2},
  arrow/.style={-Latex, thick}
]

% Shared Retrieval Agent
\node[agent, fill=yellow!20, minimum width=5.2cm] (shared) at (0, 0) {Dense Retrieval Agent};

% Grammar Extraction Crew (bottom left, closer)
\node[crew, anchor=north east] (crew1) at (-2.0, -1.7) {
  \begin{tikzpicture}[node distance=0.25cm]
    \node[agent] (a1) {Grammar Extraction Agent};
    \node[agent, below=of a1] (a2) {Grammar Formatting Agent};
    \node[tool, right=0.7cm of a2] (t1) {\texttt{Grammar Formatting Tool}};
    \draw[arrow] (a2) -- (t1);
  \end{tikzpicture}
};
\node[anchor=south west] at (crew1.north west) {\scriptsize\bfseries Grammar Extraction Crew};

% Coverage Plateau Crew (bottom right, now here)
\node[crew, anchor=north west] (crew2) at (-0.1, -1.11) {
  \begin{tikzpicture}[node distance=0.25cm]
    \node[agent] (c1) {Analysis Agent};
    \node[agent, below=of c1] (c2) {Vulnerabilities Agent};
    \node[agent, below=of c2] (c3) {Coverage Surpassing Agent};
    \node[tool, right=0.7cm of c2] (t3a) {\texttt{CVEs Retrieval Tool}};
    \node[tool, right=0.7cm of c3] (t3b) {\texttt{Packet Parsing Tool}};
    \draw[arrow] (c2) -- (t3a);
    \draw[arrow] (c3) -- (t3b);
  \end{tikzpicture}
};
\node[anchor=south east] at (crew2.north east) {\scriptsize\bfseries Coverage Plateau Crew};

% Seed Enrichment Crew (top center, now here)
\node[crew, anchor=south] (crew3) at (0, 1.11) {
  \begin{tikzpicture}[node distance=0.25cm]
    \node[agent] (b1) {Seeds Enricher Agent};
    \node[tool, right=0.7cm of b1] (t2) {\texttt{Seeds Parsing Tool}};
    \draw[arrow] (b1) -- (t2);
  \end{tikzpicture}
};
\node[anchor=south] at (crew3.north) {\scriptsize\bfseries Seed Enrichment Crew};

% Arrows from shared agent
\draw[arrow] (shared.north) -- ++(0,0.7); % up
\draw[arrow] (shared.south west) -- ++(-1.3,-1.3); % down left
\draw[arrow] (shared.south east) -- ++(1.3,-0.7); % down right

\end{tikzpicture}
\caption{{\centering \small Fig. 4. Summary of the \textit{MultiFuzz}'s crews, showing each crew’s internal agents and integrated tools. All three crews share a Dense Retrieval Agent for semantic context fetching.}}
\label{fig:crew_flow_diagram}
\end{figure*}
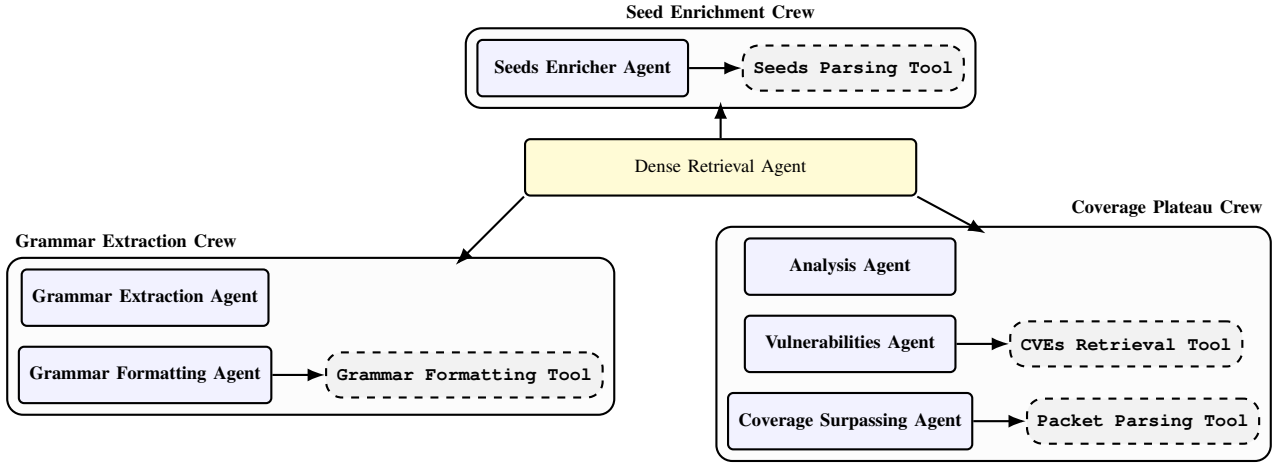

\subsection{Agentic Chunking Module}
Following the propositional transformation step, we employ an intelligent chunking mechanism termed the \textit{Agentic Chunker} to group semantically similar propositions into cohesive and operationally meaningful units. This process creates the foundation for precise retrieval and role-specific agent inference in later stages of the \textit{MultiFuzz} framework.

Formally, given a set of propositions $\mathbb{P} = \{p_1, p_2, \dots, p_n\}$ derived from the RFC document, the goal is to partition $\mathbb{P}$ into a set of non-overlapping semantic chunks $\mathcal{Z} = \{z_1, z_2, \dots, z_m\}$, where each $z_j \subseteq \mathbb{P}$ and $\bigcup_{j=1}^{m} z_j = \mathbb{P}$. The chunking objective can be viewed as an unsupervised grouping problem constrained by topic cohesion, guided by an LLM.

The chunking process is agentic in nature: each incoming proposition $p_i$ is evaluated using a prompt-driven LLM flow by \textit{gpt-4o-mini}. This LLM first examines the current set of chunk summaries and determines whether $p_i$ semantically aligns with any existing chunk $z_j$. If alignment is detected, $p_i$ is appended to that chunk. Otherwise, a new chunk is instantiated.

Each chunk $z_j$ maintains three evolving elements:
\begin{itemize}
    \item A list of constituent propositions $\{p_k\}_{k=1}^{K}$,
    \item A concise chunk summary $s_j$ generated by the LLM.
    \item A descriptive, technically precise chunk title $t_j$.
\end{itemize}

The internal logic can be modeled as a two-stage LLM pipeline:
\begin{enumerate}
    \item \textbf{Chunk Selection:} Given current chunk outlines and a new proposition $p_i$, select the most semantically compatible chunk $z_j$ such that:
    \[
    z_j = \arg\max_{z \in \mathcal{Z}} \text{sim}(p_i, s_z)
    \]
    If $\max \text{sim} < \theta$, where $\theta$ is a system-defined compatibility threshold, a new chunk is created.
    \item \textbf{Metadata Refinement:} After assignment, the system regenerates the chunk's summary $s_j$ and title $t_j$ using structured prompt templates conditioned on the current list of propositions.
\end{enumerate}

The final output is a collection of richly annotated document objects, each encapsulating a semantic group of RFC-derived propositions, along with human-readable summaries and titles. These document objects were then embedded using OpenAI's \textit{text-embedding-ada-002} model and indexed into a \textit{Chroma}-based dense vector database. A sample document object can be shown in Figure 3.

\subsection{Dense Retrieval Agent}
The first common agent in all crews is the dense retrieval agent. This agent is responsible for querying a \textit{Chroma}-based dense vector store populated with semantically grouped RTSP agentic chunks. It utilizes a \texttt{Custom RAG Tool}, to perform approximate nearest neighbor search against the indexed chunks. The output of this agent is a context-rich corpus of relevant documents passed to assist all agents with their tasks.

\subsection{Grammar Extraction Crew of Agents}
The \textit{Grammar Extraction Crew} is the first crew of agents designed to extract structured RTSP client request templates for ChatAFL. It operates through three agents: a dense retrieval agent, a grammar extraction agent, and a grammar formatting agent.

\paragraph*{\textbf{Grammar Extraction Agent}}  
This agent uses the retrieved context to produce JSON-formatted RTSP request templates, where each method (e.g. PLAY, DESCRIBE) maps to a list of headers containing \texttt{<<VALUE>>} placeholders and \texttt{\textbackslash r\textbackslash n} terminators.

\paragraph*{\textbf{Grammar Formatting Agent}}  
It refines the raw JSON output into a clean, numbered textual format using the \texttt{Grammar Extraction Formatting Tool}, making it easier to parse in the ChatAFL grammar parsing module.

% \paragraph*{\textbf{Workflow}}  
% Coordinated by the \textit{MultiFuzz} framework, the agents sequentially process context, generate templates, and produce logs. This pipeline ensures the fuzzer receives valid, context-aware structured grammar representations for improved protocol coverage.

\subsection{Seed Enrichment Crew of Agents}
The \textit{Seed Enrichment Crew} is a two-agent crew designed to enhance a given sequence of RTSP client requests by inserting new protocol-compliant packets at semantically correct positions. This enrichment supports fuzzers by generating deeper, more state-aware input sequences. The first agent is the dense retrieval agent, and the second is the seeds enricher agent.

\paragraph*{\textbf{Seeds Enricher Agent}}  
This agent interprets the protocol’s FSM and uses retrieved context from the dense retrieval agent to insert two desired client requests, typically absent from the original seed into their appropriate positions as adopted in ChatAFL. It ensures server responses are excluded and leverages the \texttt{Seeds Parsing Tool} to generate structured outputs of continuous enriched network packets. These enriched seeds are structured to be easily parsed by the ChatAFL parsing module.

% \paragraph*{\textbf{Workflow}}  
% Using the \textit{MultiFuzz} framework, the retrieval agent first collects contextual information, which is then passed to the enrichment agent. Final enriched seeds are logged in structured files, enabling reproducible and explainable seed generation. This process increases protocol state space coverage and helps fuzzers explore deeper execution paths.

\subsection{Coverage Plateau Surpassing Crew of Agents}
The \textit{Coverage Plateau Surpassing Crew} is designed to help the fuzzer escape stagnation points during test execution, where no new protocol states or code paths are being explored as observed in ChatAFL. This crew of agents aims to generate packets that can trigger new transitions by analyzing communication history, retrieved context, and optionally exploiting known CVEs.

\paragraph*{\textbf{Analysis Agent}}  
This agent performs deep context analysis of the context retrieved from the dense retrieval agent and the fuzzer’s communication history to construct a detailed generation prompt. Rather than producing packets directly, it crafts precise instructions to guide the next agent in generating a coverage-enhancing input.

\paragraph*{\textbf{Vulnerabilities Agent}}  
To improve the chance of producing impactful packets, this agent enriches the generation prompt with insights from real CVEs, fetched using a \texttt{CVEs Retrieval Tool} which uses the NVD (National Vulnerability Database) API to obtain the Live555 server vulnerabilities \cite{nvd}. If any vulnerability discovered is relevant to the current communication context, the prompt is refined accordingly; otherwise, it is forwarded unchanged.

\paragraph*{\textbf{Coverage Surpassing Agent}}  
Finally, this agent consumes the refined prompt and generates a valid RTSP client request designed to surpass the coverage plateau. The agent uses a \texttt{Packet Parsing Tool} to structure the final packet and log it along with an explanation of its purpose. A generated sample prompt can be shown in Figure 5.

\vspace{0.05cm}

{\scriptsize
\begin{tcolorbox}[
  enhanced,
  sharp corners=south,
  colframe=black!50!black,
  colback=gray!10,
  coltitle=white,
  boxrule=0.6pt,
  titlerule=0.8pt,
  title=Sample Prompt to generate Coverage Plateau Packet,
  fonttitle=\bfseries,
  fontupper=\fontsize{8}{10}\selectfont,
  label={box:cov_plateau_prompt},
]
% \hspace{-0.35cm}\{

\textbf{"prompt"}: \{ \\
"To surpass the current coverage plateau, Generate a PAUSE request that will transition the server from the Playing state to the Ready state. The PAUSE method should be sent with the appropriate headers, including CSeq: 5, Session: 000022B8, and the method set to PAUSE. This will explore the state transition from Playing to Ready, potentially revealing new server behaviors and increasing coverage." \\
\}
\end{tcolorbox}
}
\begin{center}
    \small Fig. 5. Sample prompt asking the final agent to generate a coverage plateau surpassing packet \vspace{0.025cm}
\end{center}

% \paragraph*{\textbf{Workflow}}  
% The \textit{MultiFuzz} framework orchestrates this workflow, ensuring traceability across agents through structured logs and intermediate outputs. The result is a targeted RTSP request capable of exercising new states and increasing the effectiveness of fuzz testing.

\subsection{Implementation}
We have developed \textit{MultiFuzz} on top of two agentic-AI frameworks: \textit{LangChain} \cite{langchain} and \textit{CrewAI} \cite{crewai}. \textit{LangChain} provides abstractions for building applications on top of LLMs. We use the \textit{LangChain} framework combined with the \textit{Chroma} vector store for embedding and indexing. We utilize it's features specifically in the RFC processing stage for RFC document agentic chunking and during the dense retrieval inference in all agent tasks. \textit{CrewAI}, on the other hand, is used to build autonomous multi-agent systems and provides modular assignment of agents to specific and unique roles. It supports integration with multiple LLM API providers and offers native support for tool-assisted workflows, enabling agents to interact with file systems and vector databases. We use \textit{CrewAI} in defining and orchestrating the agents that compose the \textit{MultiFuzz} framework. These structured agent groups, or "crews", coordinate within the ChatAFL framework using event-driven task scheduling augmented with custom structured tools.
% ------------- END Methodology --------------------

% ------------- START Experimental Design --------------------
\section{Experimental Design and Evaluation}\label{ex-des}
We evaluate the proposed \textit{MultiFuzz} framework by measuring its effectiveness in fuzzing stateful protocol implementations using a multi-agent-based architecture. Our evaluation aims to answer the following research questions:

\begin{itemize}
    \item \textbf{RQ1:} How effective is \textit{MultiFuzz} in improving branch coverage and state exploration compared to SOTA protocol fuzzers?
    \item \textbf{RQ2:} How does the multi-agent collaboration strategy improve over single-LLM approaches such as ChatAFL?
\end{itemize}

To conduct the evaluation, we test \textit{MultiFuzz} on the RTSP protocol implemented by the Live555 media streaming server. RTSP was selected due to its rich stateful behavior, complexity in session semantics, and widespread use in multimedia transmission. It presents non-trivial state transitions that make it a fitting candidate for state-aware fuzzing. The framework is powered by Llama-based language models obtained via the Groq-Cloud API \cite{groq}, which are: \textit{llama3.3-70b-versatile}, \textit{deepseek-r1-distill-llama-70b}, \textit{llama3-70b-8192}, \textit{llama-4-scout-17b-16e-instruct}, and \textit{llama-3.1-8b-instant}, chosen for their reasoning abilities, and long-context window capacities. Throughout the experimentation process, we explored different combinations of these models across the various agent groups in the framework. Tasks such as grammar extraction, seed enrichment, and plateau surpassing were assigned to different models iteratively until the most effective model was identified for each specific subtask, optimizing the overall performance of \textit{MultiFuzz}.

\subsection{Experiments Setup}
To evaluate the fuzzing effectiveness of \textit{MultiFuzz}, we conducted a 24-hour three fuzzing sessions using our framework alongside the three SOTA baseline fuzzers: NSFuzz, AFLNet, and ChatAFL under the same experimental conditions. All experiments were performed on a local machine running Ubuntu 24.04.02 LTS, equipped with an Intel Core i5-11300H processor and 16 GB of RAM. The fuzzers were evaluated against the RTSP protocol implemented in the Live555 media streaming server. Each fuzzer was independently executed with default settings. We measured the effectiveness of each fuzzer across several key metrics, including unique crashes, state coverage, branch coverage, and total paths explored. This setup allows us to assess the relative performance of \textit{MultiFuzz} in contrast with existing approaches.

\subsection{Evaluation Metrics}
To evaluate \textit{MultiFuzz}'s fuzzing performance, we adopted standard coverage-based metrics inspired by existing works such as AFLNet and ChatAFL:

\begin{itemize}
    % \item \textbf{Unique Crashes:} Number of distinct inputs that triggered a crash on the target server under test (SUT).
    \item \textbf{Branch Coverage:} Number of unique conditional branches exercised in the code.
    \item \textbf{Number of States:} Number of FSM states reached and explored during fuzzing.
    \item \textbf{Number of State Transitions:} The total count of valid state transitions triggered within the protocol’s FSM, reflecting the depth of state space exploration.
\end{itemize}

We use ProFuzzBench \cite{profuzzbench} as our benchmarking platform due to its automated nature in a containerized environment using Docker and to baseline with the previous SOTA protocol fuzzers. All experiments were repeated multiple times to ensure consistency, and results were averaged over time windows to account for variability in execution. This setup allows us to rigorously measure \textit{MultiFuzz}'s capability to intelligently generate protocol-aware inputs and uncover deep-state vulnerabilities.
% ------------- END Experimental Design --------------------

% ------------- START Experimental Results --------------------

\subsection{Experimental Results of Fuzzing on Code Coverage}
Table I presents the branch coverage results, demonstrating \textit{MultiFuzz}'s substantial superiority in code coverage metrics. \textit{MultiFuzz} achieves an average branch coverage of 2940 branches, representing dramatic improvements of 1.0\% over ChatAFL (2912.67), 2.8\% over AFLNet (2860), and 2.3\% over NSFuzz (2807). Although these percentage improvements may appear modest, the absolute differences are significant in the context of protocol fuzzing, where each additional branch represents potential discovery of critical vulnerabilities. The consistency of \textit{MultiFuzz}'s performance is particularly noteworthy, with coverage ranging from 2970 to 2940 branches across experiments, demonstrating reliable and predictable performance. In contrast, ChatAFL shows higher variability (2890-2998 branches), while AFLNet exhibits perfect consistency but at significantly lower coverage levels. NSFuzz demonstrates the most variability, with coverage ranging from 2795 to 2826 branches.

\section{Experimental Results and Discussion}\label{res}
\subsection{Experimental Results of Fuzzing on State Exploration}
Table II and Table III demonstrate that compared to NSFuzz, AFLNet, and ChatAFL, \textit{MultiFuzz} achieves superior performance in both state transitions and state exploration across all three experimental runs. In terms of state transitions, \textit{MultiFuzz} achieves an average of 163.33 transitions, representing a significant improvement of 2.3\% over ChatAFL (159.67), 94.5\% over AFLNet (84.0), and 81.2\% over NSFuzz (90.33). The state exploration results further validate \textit{MultiFuzz}'s effectiveness, with an average of 14.67 states explored compared to ChatAFL's 14.33 states, AFLNet's 10.0 states, and NSFuzz's 11.7 states. This represents improvements of 2.4\%, 46.7\%, and 25.4\% respectively.

\begin{table}[htbp]
% \centering
% \caption{\centering \small Branch coverage achieved by \textit{MultiFuzz} and baseline SOTA fuzzers}
\begin{center}
\small \textbf{TABLE I}: Branch coverage achieved by \textit{MultiFuzz} and baseline SOTA fuzzers
\end{center}
\label{tab:branch_coverage}

\begin{adjustbox}{width=\columnwidth}
\begin{tabular}{|c|c|c|c|c|}
\hline
\textbf{Experiment} & \textbf{MultiFuzz} & \textbf{ChatAFL} & \textbf{AFLNet} & \textbf{NSFuzz} \\
\hline
1 & 2970 & 2890 $\uparrow$ 2.8\% & 2850 $\uparrow$ 4.2\% & 2800 $\uparrow$ 6.1\% \\
\hline
2 & 2910 & 2998 $\downarrow$ -2.9\% & 2870 $\uparrow$ 1.4\% & 2795 $\uparrow$ 4.1\% \\
\hline
3 & 2940 & 2850 $\uparrow$ 3.2\% & 2860 $\uparrow$ 2.8\% & 2826 $\uparrow$ 4.0\% \\
\hline
\textbf{Average} & \textbf{2940.0} & \textbf{2912.67 $\uparrow$ 0.9\%} & \textbf{2860.0 $\uparrow$ 2.8\%} & \textbf{2807.0 $\uparrow$ 4.7\%} \\
\hline
\end{tabular}
\end{adjustbox}
\end{table}

\begin{table}[htbp]
% \centering
% \caption{\centering \small Number of state transitions achieved by \textit{MultiFuzz} and baseline SOTA fuzzers}
\begin{center}
\small \textbf{TABLE II}: Number of state transitions achieved by \textit{MultiFuzz} and baseline SOTA fuzzers
\end{center}
\label{tab:state_transitions}
\begin{adjustbox}{width=\columnwidth}
\begin{tabular}{|c|c|c|c|c|}
\hline
\textbf{Experiment} & \textbf{MultiFuzz} & \textbf{ChatAFL} & \textbf{AFLNet} & \textbf{NSFuzz} \\
\hline
1 & 163 & 159 $\uparrow$ 2.5\% & 80 $\uparrow$ 103.8\% & 88 $\uparrow$ 85.2\% \\
\hline
2 & 165 & 158 $\uparrow$ 4.4\% & 85 $\uparrow$ 94.1\% & 91 $\uparrow$ 81.3\% \\
\hline
3 & 162 & 162 $\uparrow$ 0.0\% & 87 $\uparrow$ 86.2\% & 92 $\uparrow$ 76.1\% \\
\hline
\textbf{Average} & \textbf{163.33} & \textbf{159.67 $\uparrow$ 2.3\%} & \textbf{84.0 $\uparrow$ 94.4\%} & \textbf{90.33 $\uparrow$ 80.8\%} \\
\hline
\end{tabular}
\end{adjustbox}
\end{table}

\begin{table}[htbp]
% \centering
% \caption{\centering \small Number of states explored by \textit{MultiFuzz} and baseline SOTA fuzzers}
\begin{center}
\small \textbf{TABLE III}: Number of states explored by \textit{MultiFuzz} and baseline SOTA fuzzers
\end{center}
\label{tab:states_explored}
\begin{adjustbox}{width=\columnwidth}
\begin{tabular}{|c|c|c|c|c|}
\hline
\textbf{Experiment} & \textbf{MultiFuzz} & \textbf{ChatAFL} & \textbf{AFLNet} & \textbf{NSFuzz} \\
\hline
1 & 14 & 14 $\uparrow$ 0.0\% & 9 $\uparrow$ 55.6\% & 12 $\uparrow$ 16.7\% \\
\hline
2 & 15 & 14 $\uparrow$ 7.1\% & 11 $\uparrow$ 36.4\% & 11 $\uparrow$ 36.4\% \\
\hline
3 & 15 & 15 $\uparrow$ 0.0\% & 10 $\uparrow$ 50.0\% & 12 $\uparrow$ 25.0\% \\
\hline
\textbf{Average} & \textbf{14.67} & \textbf{14.33 $\uparrow$ 2.4\%} & \textbf{10.0 $\uparrow$ 46.7\%} & \textbf{11.7 $\uparrow$ 25.4\%} \\
\hline
\end{tabular}
\end{adjustbox}
\end{table}

\subsection{Observations}
The dense retrieval-based multi-agent system of \textit{MultiFuzz} enables more systematic state space exploration by leveraging the agented-chunked and embedded protocol specifications and coordinated agent interactions. Unlike baseline fuzzers that rely on conventional feedback-driven exploration, \textit{MultiFuzz}'s multi-agent architecture facilitates comprehensive state discovery through intelligent coordination and knowledge sharing among specialized agents. The results indicate that \textit{MultiFuzz}'s dense retrieval mechanism effectively identifies and prioritizes valuable states that serve as critical transition points within the protocol state machine. The multi-agent coordination allows for parallel exploration strategies while maintaining systematic coverage of the state space. This approach significantly outperforms traditional fuzzing methods that rely on random mutation and single LLM approaches.

% ------------- START Conclusion --------------------
\section{Conclusion and Future Work}\label{conc}
Protocol fuzzing continues to be a foundational technique for uncovering implementation flaws in communication systems. However, traditional fuzzers often face significant limitations when applied to stateful or proprietary network protocols, particularly due to difficulties in handling complex message grammars, managing multi-step state transitions, and maintaining valid interactions across sessions. We proposed \textit{MultiFuzz}, a dense retrieval-based multi-agent system designed to address these limitations by leveraging an agentic-RAG-based architecture empowered by chain-of-thought reasoning. Our approach builds upon prior advances in LLM-assisted fuzzing but distinguishes itself by introducing multi-agent coordination instead of a single LLM. It proposes agentic-based chunking of protocol documents and context-aware inference about protocol specifications and vulnerabilities. These additions help overcome key challenges in stateful fuzzing, such as low coverage and stagnation during long-running sessions. Evaluation across real-world protocol implementations has shown that \textit{MultiFuzz} surpasses existing tools such as NSFuzz, AFLNet, and ChatAFL in terms of state exploration and branch coverage. These findings bridge the gap between traditional fuzzing methodologies and recent advances in agentic-AI, as they open promising opportunities for more effective and adaptive security testing.

Looking forward, we suggest several paths to extend this work. Firstly, enhancing automation by tightly integrating reverse engineering tools, symbolic analyzers, and traffic parsers can further streamline the entire pipeline. Lastly, fine-tuning agent behaviors using domain-specific interaction data could improve their effectiveness in specialized protocol domains.

\section{Acknowledgment}
Heartfelt gratitude is extended to AiTech AU, \textit{AiTech for Artificial Intelligence and Software Development} (\url{https://aitech.net.au}), for funding this research and enabling its successful completion.

% In summary, \textit{MultiFuzz} represents a step toward intelligent, protocol-aware fuzzing powered by collaborative agents and modern LLM infrastructure. By bridging the gap between traditional fuzzing methodologies and recent advances in generative AI, it opens promising opportunities for more effective and adaptive security testing.
% ------------- END Conclusion --------------------

% \section{Acknowledgment}
% Heartfelt gratitude is extended to AiTech AU, \textit{AiTech for Artificial Intelligence and Software Development} (\url{https://aitech.net.au}), for funding this research, and enabling its successful completion.

\vspace{-0.2cm}

\bibliographystyle{IEEEtran}
\bibliography{references}
\end{document}